\documentstyle[12pt]{article}
\newcommand{\sptwo}{1.4}

\newcommand{\doublespace}{\edef\baselinestretch{\sptwo}\Large\normalsize}

\begin{document}
\doublespace

\begin{center}
{\bf Equivalent Linear Two-Body Equations for Many-Body Systems
}\\
Alexander L. Zubarev and Yeong E. Kim\\
Department of Physics, Purdue University\\
West Lafayette, Indiana  47907\\
{\bf Abstract}
\end{center}
\begin{quote}
A method has been developed for obtaining equivalent linear two-body 
equations (ELTBE)
for the system of many ($N$) bosons using the variational
principle. The method has been applied to the
one-dimensional $N$-body problem with pair-wise contact
interactions (McGurie-Yang $N$-body problem)
and to the dilute Bose-Einstein condensation (BEC) of
atoms in anisotropic harmonic traps at zero temperature. For
both cases, it is
shown that the method gives excellent results for large $N$.
\end{quote}

\noindent
PACS numbers:  03.75.Fi, 03.65.Db, 05.30.Jp, 67.90.+z

\pagebreak

In this paper we present an approximate method of obtaining the
eigenvalue solutions of the system of interacting $N$ bosons
using an equivalent two-body method similar to that used by Feshbach
and Rubinov [1] for the triton ($^3H$) three-body ($N$=3) bound state.
 They [1] used both (i) the variational principle
and (ii) a reduced coordinate variable ( not the hyperradius)
to obtain an equivalent two-body equation for the three-body
bound state ($^3H$).
For many-body problems, use
of one 
reduced coordinate variable
 (the hyperradius [2])
was made to obtain equivalent two-body equations
by keeping only a finite sum of terms  of
the hyperspherical expansion with $K=K_{min}$
($K$ is the global angular momentum). This
method
has been applied to the ground state of the $N$-body 
system composed of distinguishable particles or of bosons
and also to nuclear bound states [3,4]. It was shown that the
method leads to the correctly behaved nuclear bound states in the
limit of large $A$ ($A$ is the nucleon number) [4]. 
Recently, it has been
used to describe the Bose-Einstein condensation (BEC) of atoms 
 in isotropic harmonic traps [5].

In this paper we apply our method to the one-dimensional $N$-body
problem with pair-wise contact interactions (the McGuire-Yang $N$-body
problem [6, 7]) and to the dilute BEC of atoms in anisotropic harmonic
traps at zero temperature. We show that the method gives
excellent results for large $N$ for both cases.  

For the $N$-body system, our method for obtaining 
 the equivalent linear two-body equation (ELTBE) consists of two
steps. The first step
is to give the $N$-body wave function $\Psi({\bf r_1},
{\bf r_2}, ...)$ a particular functional form
$$
\Psi({\bf r_1},{\bf r_2}, ...) \approx \tilde{\Psi}(\zeta_1,
\zeta_2, \zeta_3),
\eqno{(1})
$$
where $\zeta_1, \zeta_2, $and $\zeta_3$ are known functions. 
We limit $\zeta$'s to three variables in order to obtain the
ELTBE.
 The second
step is to
derive an equation for $ \tilde{\Psi}(\zeta_1, \zeta_2,
\zeta_3)$  by requiring that $
\tilde{\Psi}$ must satisfy a variational principle
$$
\delta \int \tilde{\Psi}^\ast \tilde{\Psi}d \tau = 0
\eqno{(2)}
$$
with a subsidiary condition $\int \tilde{\Psi}^\ast \tilde{\Psi}d
\tau =1$.
This leads to a linear two-body equation, from  which both
eigenvalues and eigenfunctions can be obtained.
The lowest
eigenvalue is an upper bound of the lowest eigenvalue of the
original N-body problem.
To test our ELTBE method, we apply the method to the
one-dimensional $N$-body problem and to the BEC of atoms in 
anisotropic traps at zero temperature in the following.

There are only several known cases of exactly solvable three-body and
four-body problems. For $N=3$ case it was shown [8] that
the Faddeev equations [9] for one-dimensional three-body
problem with pair-wise contact interactions are exactly solvable.
For the one-dimensional  $N=4$ case, analytical solutions 
of the four-body
Faddeev-Yakubovsky were obtained in [10].
We note that for nuclear three body systems with short-range 
interactions, the  Schr\"odinger equation in three-dimension 
is reformulated into
the Faddeev equations [9] which can be solved numerically after
making
partial wave expansion [11] or without partial wave expansion
[12]. 
In the following, we consider an exactly solvable one-dimensional
$N$-body system as a test case for our method.

For the one-dimensional N-body problem with the Hamiltonian
$H=-\sum_{i=1}^{i=N} d^2/dx_{i}^2+2c \sum_{i<j} \delta(x_i-x_j)$
(with  $\hbar=m=1$),
the Schr\"odinger equation is exactly solvable. The bound and
scattering states for this system have been found by McGuire [6]
and by Yang [7].
For the case $c<0$, there are bound states [6] for the system of
$N$ bosons with the wave function $\Psi= \exp[(c/2) \sum_{i<j} \mid
x_i-x_j \mid]$. The energy of this bound state is 
$$ E_{exact}=-c^2 N(N^2-1)/12.
\eqno{(3)}
$$

The McGuire-Yang N-body problem gives a unique possibility to
check the validity of various approximations made for the
Schr\"odinger equation describing $N$ particles interacting via short
range potential. 

For this case, we seek for eigenfunction $\Psi$ of $H$ in the form of
$$  
\Psi \approx \tilde{\Psi}( \rho) = \frac{F( \rho)}{\rho^{(N-2)/2}}, 
$$
where $ \rho= \frac{1}{N} \sum_{i<j} (x_i-x_j)^2$. 
We shall derive an equation for $ F( \rho )$ by requiring that $
\tilde{\Psi}$ must satisfy a variational principle (2).
This requirement leads to the equation
$$ [- \frac{d^2}{d \rho^2}+\frac{(N-2)(N-4)}{4 \rho^2}+V( \rho)]
F( \rho) = EF( \rho),
\eqno{(4)}
$$
where
$$
V( \rho) = cN(N-1) \frac{ \Gamma (N/2-1/2)}{\sqrt{2 \pi } \Gamma
(N/2-1)} \frac{1}{ \rho}.
\eqno{(5)}
$$

 We note that Eq. (4) is exactly the form of the
Schr\"odinger two-body equation in which $ \Psi=(F( \rho)/
\rho)Y_{lm}$, and  a centrifugal potential energy is given by $
(N-2)(N-4)/(4 \rho^2r)$ with identification of angular momentum
quantum number
$l=N/2-1$. Eq. (4) with the Coulomb like potential $V( \rho)$,
Eq. (5), can be solved analytically, and we obtain for E the
following expression [13]
$$E=-\frac{c^2}{2 \pi}[ \frac{N(N-1)
\Gamma(N/2-1/2)}{(N-2) \Gamma(N/2-1)}]^2.
\eqno{(6)}
$$
In the case of large $N$, using the asymptotic formulas for
$\Gamma$ function,
$
\lim_{\mid z\mid \rightarrow \infty} \frac{\Gamma(z+\alpha)}
{\Gamma(z+\beta)}=z^{\alpha-\beta}(1+O(\frac{1}{z}))
$,
we obtain $E=-\frac{c^2}{4 \pi}N^3$ for the leading term of Eq. (6).
 On the other hand we have for large $N$ case from Eq. (1), $E_{exact}=
-\frac{c^2}{12}N^3(1+O(\frac{1}{N^2}))$.
Therefore, for the McGuire-Yang N-body problem
we have demonstrated that the 
ELTBE method, Eqs. (4) and (5), is a very good
approximation for the case of large $N$
(the relative error for binding energy is about 4.5 \%).
Furthermore, our approximation, Eq. (6), agrees remarkably well
with exact value, Eq. (3), for any $N$ (the maximum value of binding
energy relative error occurs for $N=3$ and is about 10 \%).

Now, let us consider $N$ identical bosonic atoms confined in a
harmonic anisotropic trap with the following Hamiltonian

$$
H=-\frac{\hbar^2}{2m} \sum_{i=1}^{N} \Delta_{i}+\frac{1}{2}
\sum_{i=1}^{N}m(\omega_x^2x_i^2+\omega_y^2y_i^2+\omega_z^2z_i^2)
+\sum_{i<j}V_{int}({\bf r}_i-{\bf r}_j),
\eqno{(7)}
$$
where  we assume $V_{int}$ in the dilute condensate case to be the
following form [14]
$$
V_{int}({\bf r}_i-{\bf r}_j)=\frac{4 \pi \hbar^2a}{m} \delta({\bf
r}_i-{\bf r}_j),
$$
 with $s$-wave scattering length, $a$.

For eigenfunction $\Psi$ of $H$, we assume the solution for
$\Psi$ has the following form
$$
\Psi({\bf r}_1, ...{\bf r}_N) \approx \frac{ \tilde{\Psi}(x,y,z)}{
(xyz)^{(N-1)/2}}
\eqno{(8)}
$$
where $x^2=\sum_{i=1}^{N}x_i^2$, $~y^2=\sum_{i=1}^{N}y_i^2$,
$~
z^2=\sum_{i=1}^{N}z_i^2$.

We now derive an equation for $\tilde{\Psi}(x,y,z)$ by requiring
that $\tilde{\Psi}(x,y,z)$ must satisfy the variational principle
(2). This requirement leads to the equation
$$
H \tilde{\Psi}=E\tilde{\Psi},
\eqno{(9)}
$$
where
$$
H=-\frac{\hbar^2}{2m}(\frac{\partial^2}{\partial x^2}+\frac{\partial^2}
{\partial y^2}+\frac{\partial^2}{\partial z^2})+\frac{m}{2}(\omega_x^2x^2
+\omega_yy^2+\omega_z^2z^2)+\frac{\hbar^2}{2m}\frac{(N-1)(N-3)}{4}(
\frac{1}{x^2}+\frac{1}{y^2}+\frac{1}{z^2})+\frac{g}{xyz},
\eqno{(10)}
$$
with $g=g_0(2 \pi)^{-3/2}(\Gamma(N/2)/ \Gamma(N/2-1/2))^3N(N-1)/2$
and $~
g_0=4 \pi \hbar^2 a/m$. To the best of our knowledge,
Eqs. (9) and (10) have not been discussed in the literature.

 For the positive scattering length case, $a>0$, we
look for the solution of Eq. (9) of the form
$$
\tilde{\Psi}(x,y,z)=\sum_{i,j,k}c_{ijk} \Phi_i^{(1)}(x)
\Phi_j^{(2)}(y) \Phi_k^{(3)}(z),
\eqno{(11)}
$$
 where $\Phi_i^{(1)}(x)=x^{(N-1)/2} \exp[-(x/ \alpha_i)^2/2]$,
$\Phi_j^{(2)}(y)=y^{(N-1)/2} \exp[-(y/ \beta_j)^2/2]$,
$\Phi_k^{(3)}(z)=z^{(N-1)/2} \exp[-(z/ \gamma_k)^2/2]$,
and $c_{ijk}$ are solutions of the following equations
$$
\sum_{l,m,n}H_{ijk,lmn}c_{lmn}=E \sum_{l,m,n} \lambda_{ijk,lmn}c_{lmn}
\eqno{(12)}
$$
where 
$$ 
H_{ijk,lmn}=\frac{\hbar
\tilde{\omega}N\lambda_{ijk,lmn}}{2}[\frac{1+\alpha_i^2
\alpha_l^2 \alpha_x^2}{\alpha_i^2+\alpha_l^2}+ \frac{1+\beta_j^2
\beta_m^2 \alpha_y^2}{\beta_j^2+\beta_m^2}+ \frac{1+\gamma_k^2
\gamma_n^2 \alpha_z^2}{\gamma_k^2+\gamma_n^2}+ 
\tilde{g}
\frac{ \sqrt{(\alpha_i^2+\alpha_l^2)(\beta_j^2+\beta_m^2)(\gamma_k^2+\gamma_n^2)}}{\alpha_i \alpha_l \beta_j \beta_m \gamma_k \gamma_n}],
$$

$$
\lambda_{ijk,lmn}=[\frac{8 \alpha_i \alpha_l \beta_j \beta_m
\gamma_k \gamma_n}{(\alpha_i^2+\alpha_l^2)(\beta_j^2+\beta_m^2)
(\gamma_k^2+\gamma_n^2)}]^{N/2},
$$
 with $\tilde{g}=\frac{(N-1)}{2 \sqrt{2}N}\tilde{n}$, $\tilde{n}=
2\sqrt{\tilde{\omega}m/(2
\pi \hbar)}Na$, $\tilde{\omega}=(\omega_x \omega_y \omega_z)^{\frac{1}{3}}
$, $\alpha_x=\omega_x/\tilde{\omega}$,
$\alpha_y=\omega_y/\tilde{\omega}$, and
$\alpha_z=\omega_z/\tilde{\omega}$.

For the case of large $N$, we have $\lambda_{ijk,lmn} \approx \delta_{il}
\delta_{jm} \delta_{kn}$, $H_{ijk,lmn} \approx E \delta_{il}
\delta_{jm} \delta_{kn}$.
For the ground state energy, using $\frac{\partial E}{\partial
\alpha_i}=\frac{\partial E}{\partial\alpha_j}=\frac{\partial E}
{\partial\alpha_k}=0$, we obtain
$$
\frac{E}{N \hbar \tilde{\omega}}=\frac{5}{4}\tilde{n}^{\frac{2}{5}}
\eqno{(13)}
$$
We note that Eq. (13) is the exact ground state solution of Eq. (9)
for large $N$. For the case of large $N$ we can obtain an
essentially exact expression for the ground state energy
by neglecting the kinetic energy term in the
Ginzburg-Pitaevskii-Gross (GPG) equation [16]
 (the Thomas-Fermi
approximation [15]) as
$$
\frac{E_{TF}}{N \hbar \tilde{\omega}}=\frac{5}{7}(\frac{15}{8}
\sqrt{ \pi})^{\frac{2}{5}} \tilde{n}^{\frac{2}{5}}
\eqno{(14)}
$$

Comparising Eq. (13) with Eq. (14), we can see that for the case of 
large $N$ the
ELTBE method (Eqs. (9) and Eq. (10)) is a very good approximation, 
with a relative 
error of about 8$\%$
for the binding energy. 

When the scattering length is negative, the effective
interaction between atoms is attractive. It has been claimed that
the BEC in free space is impossible [17] because the attraction
makes the system tend to an ever dense phase. For $^7Li$, the
$s$-wave scattering length is $a=(-14.5 \pm 0.4) \AA$ [18]. For
bosons trapped in an external potential there may exist a
metastable BEC state with a number of atoms below the critical
value $N_{cr}$ [19-27].  

For the $a<0$ case, we can see that potential energy  in Eq. (10),
$$
V(x,y,z)=\frac{m}{2}(\omega_x^2x^2+\omega_y^2y^2+\omega_z^2z^2)+
\frac{\hbar^2}{2m}\frac{(N-1)(N-3)}{4}(\frac{1}{x^2}+\frac{1}{y^2}+
\frac{1}{z^2})-
\frac{ \mid g \mid }{xyz},
\eqno{(15)}
$$
for $N<N_{cr}$ has a single metastable minimum which leads to the
metastable BEC state. We note that for the case of large $N_{cr}$,
the ELTBE method leads to the same $N_{cr}$ as the variational GPG 
stationary
theory [26]. To show this, let us consider an anisotropic trap, $
\omega_x=\omega_y=\omega_\bot$, $\omega_z=\lambda \omega_\bot $.
Local minimum conditions $\hat{A} > 0$, where $\hat{A}$ is a
matrix with matrix elements $A_{ij}=\partial^2V/ \partial x_i
\partial x_j$, can be written for this case as
$$
n^2/2 \delta_\bot^2 \delta_z^{\frac{1}{2}}-n-
\lambda ^2 \delta_\bot \delta_z^{\frac{3}{2}}/32+O(\frac{1}{N})
< 0,
\eqno{(16)}
$$
where $\delta_z=(2m \omega_\bot/\hbar N_{cr})z^2,
\delta_\bot=(2m \omega_\bot/ \hbar N_{cr})x^2$, and $n=2(m
\omega_\bot/2 \pi \hbar)^{1/2}N_{cr}\mid a \mid$.
Setting the left-hand side of Eq. (16) to zero and
neglecting $O(\frac{1}{N})$ terms, we obtain the following
equations for $N_{cr}$ 
$$ 1 - 2\delta_\bot^2 = \delta_\bot^2 (1 + 8
\frac{\delta_z}{\delta_\bot} \lambda^2)^{1/2} ,
1 - \lambda^2 \delta_z^2 = \delta_z\delta_\bot[1 + (1 + 8
\frac{\delta_z}{\delta_\bot} \lambda^2)^{1/2}],
n = \delta_z^{1/2}\delta_\bot^2[1 + (1 + 8
\frac{\delta_z}{\delta_\bot} \lambda^2)^{1/2}] .
\eqno{(17)}
$$

\noindent
Eqs. (17) are exactly the same as equations for determining $N_{cr}$
obtained from the variational GPG approach [26].
Taking the experimental values of $^7Li$ trap parameters [28], $\omega_\bot /2
\pi$ = 152 Hz, and $\omega_z/2 \pi$ = 132 Hz, we obtain $
N_{cr}$= 1456. This value of $ N_{cr}$ is consistent with
theoretical predictions [23-27] and is in agreement with those
observed in a recent experiment [28].

We note that the ELTBE method for a
general anisotropic trap can be improved using a generalization
of hyperspherical expansion
$$
\Psi({\bf r}_1, ...{\bf
r}_N)=\sum_{\stackrel{K_x,K_y,K_z,}{ \nu_x,\nu_y, \nu_z}}
\Psi_{K_x,K_y,K_z}^{  \nu_x,\nu_y, \nu_z}(x,y,z)Y^{\nu_x}_{K^{\nu_x}_x}(\Omega_x)Y^{\nu_y}_{K_y}(\Omega_y)Y^{\nu_z}_{K_z}(\Omega_z),
\eqno{(18)}
$$
where the hyperspherical harmonics $Y^{\nu_x} _{K_x}(\Omega_x)$, $
Y^{\nu_y}_{K_y}(\Omega_y)$, and $Y^{\nu_z}_{K_z}(\Omega_z)$ 
are eigenfunctions of the
angular parts of the Laplace operators $\sum_{i=1}^{N} \frac{\partial^2}
{\partial x_i^2}$, $\sum_{i=1}^{N} \frac{\partial^2}{\partial
y_i^2}$, and $\sum_{i=1}^{N} \frac{\partial^2}{\partial z_i^2}$,
respectively. However, we do not expect a fast convergence of the
expansion, Eq. (18), because of nonuniformity  
of the convergence of the expansion of
$
\sum_{i<j}V_{int}({\bf r}_i-{\bf r}_j)$ in $x$, $y$, and $z$.

In summary, we have presented a method for obtaining an 
equivalent linear two-body equation
 for the system of $N$ bosons. We have applied
the method to the McGuire-Yang N-body problem and also to the dilute
Bose-Einstein condensation in anisotropic harmonic traps at zero
temperature. For both cases we have shown that the method gives excellent results
for large $N$.
 
\pagebreak

\begin{center}
{\bf References}
\end{center}

\vspace{8pt}

\noindent
[1] H. Feshbach and S.I. Rubinov, Phys. Rev. {\bf 98}, 188 (1955).

\noindent
[2] V. A. Fock, Izv. Akad. Nauk SSSR, ser. Fiz. {\bf 18}, 161
(1954).

\noindent
[3] A. M. Badalyan, F. Calogero, Yu. A. Simonov, Nuovo Cimento
{\bf 68}A, 572 (1970);
F. Calogero, Yu. A. Simonov, Nuovo Cimento
{\bf 67}A, 641 (1970); F. Calogero et al, Nuovo Cimento {\bf
14}A, 445 (1973); F. Calogero, Yu. A. Simonov, Phys. Rev. {\bf
169}, 789 (1967); A. I. Baz' et al., Fiz. Elem. Chastits At.
Yadra {\bf 3}, 275 (1972) [Sov. J. Part. Nucl. {\bf 3},137
(1972)].

\noindent
[4] F. Calogero, Yu. A. Simonov, and E. L. Surkov, Nuovo Cimento
{\bf 1}A, 739 (1971).

\noindent
[5] J. L. Bohn, B. D. Esry, and C. H. Greene, Phys. Rev. A{\bf
58}, 584 (1998).

\noindent
[6] J. B. McGuire, J. Math. Phys. {\bf 5}, 622 (1964);
J. Math. Phys. {\bf 7}, 123 (1966).

\noindent
[7] C. N. Yang, Phys. Rev. Lett. {\bf 19}, 1312 (1967);
Phys. Rev. {\bf 168}, 1920 (1967).

\noindent
[8] L. R. Dodd, J. Math. Phys. {\bf 11}, 207 (1970);
                Phys. Rev. D{\bf 3}, 2536 (1971);
                Aust. J. Phys. {\bf 25}, 507 (1972).

\noindent
[9] L.D. Faddeev, { \it Mathematical Aspects of the Three-Body
Problem in the Quantum Scattering Theory } (Daniel Davey and
Company, Inc., New York, 1965).

\noindent
[10] A. L. Zubarev and V. B. Mandelzweig, Phys. Rev. C{\bf 52}, 509
(1995).

\noindent
[11] E.P. Harper, Y.E. Kim, and A. Tubis, Phys. Rev. Lett. {\bf
23}, 1533 (1972); Phys. Rev. C{\bf 2}, 877 (1970); Phys. Rev.
C{\bf 2}, 2455(E) (1970); Phys. Rev. C{\bf 6}, 126 (1972).

\noindent
[12] R. A. Rice and Y.E. Kim, Few Body Systems {\bf14}, 127
(1993).

\noindent
[13] Y. E. Kim and A. L. Zubarev ``{ \it Effective  Linear
Two-Body Method for Many-Body Problem}", Purdue Nuclear and
Many-Body Theory Group (PNMBTG) preprint, PNMBTG-99-2, to be
submitted to Annals of Physics (N.Y.). This reference  contains detailed 
derivations of all formulae given without derivations in this paper.

\noindent
[14] E. Fermi, Riverica Sci. {\bf7}, 13 (1936).

\noindent
[15] G. Baym and C. J. Pethick, Phys. Rev. Lett. {\bf 76}, 6
(1996).

\noindent
[16] L. Ginzburg, and L. P. Pitaevskii, Zh. Eksp. Teor. Fiz.
{\bf 34}, 1240
(1958) [Sov. Phys. JETP 7, 858 (1958)]; E. P. Gross, J. Math.
Phys. {\bf 4},
195 (1963).

\noindent
[17] T. D. Lee, K. Huang, and C. N. Yang, Phys. Rev. {\bf 106},
1135 (1957).

\noindent
[18]  E. R. I. Abraham, W. I. McAlexander, C. A. Sackett, and R.
G. Hulet,
Phys. Rev. Lett. {\bf 74}, 1315 (1995).

\noindent
[19]   Y. Kagan, G. V. Shlyapnikov, and J. T. M. Walraven, Phys.
Rev. Lett.
{\bf 76}, 2670 (1996). 

\noindent
[20]  Alexander L. Fetter, cond-mat/9510037.

\noindent
[21] M. Houbiers and H.T.C. Stoof, Phys. Rev. A {\bf 54}, 5055
(1996).

\noindent
[22] E. V. Shuryak, Phys. Rev. A {\bf 54}, 3151 (1996).

\noindent
[23] M. Ueda, and A. J. Leggett, Phys. Rev. Lett. {\bf 80}, 1576
(1998).

\noindent
[24]  F. Dalfovo and S.Stringari, Phys. Rev. {\bf A53}, 2477
(1996).

\noindent
[25]  R. J. Dodd, M. Edwards, C. J. Williams, C. W. Clark, M. J.
Holland,
P. A. Ruprecht, and K. Burnett, Phys. Rev. {\bf A54}, 661 (1996).

\noindent
[26] Y. E. Kim and A. L. Zubarev, Phys. Lett. A{\bf 246}, 389
(1998).

\noindent
[27] M. Wadati and T. Tsurumi, Phys. Lett. A{\bf 247}, 287
(1998).

\noindent
[28]   C. C. Bradley, C. A. Sackett, and R. G. Hulet, Phys. Rev.
Lett.
{\bf 78}, 985 (1997).

\end{document}